\documentclass[epj,final]{svjour}
\usepackage{graphicx,amsmath,bm}

\begin{document}

\title{Optical properties of ZnCr$_\text{2}$Se$_\text{4}$}
\subtitle{Spin-phonon coupling and electronic \textit{\rmfamily
d}-\textit{\rmfamily d}-like excitations}

\author{T.~Rudolf\inst{1}\thanks{\emph{Present address:} torsten.rudolf@physik.uni-augsburg.de}, Ch.~Kant\inst{1}, F.~Mayr\inst{1}, M.~Schmidt\inst{1}, V.~Tsurkan\inst{1,2}, J. Deisenhofer\inst{1} \and A.~Loidl\inst{1}
} \institute{Experimental Physics V, Center for Electronic Correlations and
Magnetism, University of Augsburg, D-86135~Augsburg, Germany \and Institute of
Applied Physics, Academy of Sciences of Moldova, MD-2028~Chi\c{s}in\u{a}u,
Republic of Moldova}

\date{\today}

\abstract{We studied the optical properties of antiferromagnetic ZnCr$_2$Se$_4$
by infrared spectroscopy up to 28,000~cm$^{-1}$ and for temperatures from 5 to
295~K. At the magnetic phase transition at 21~K, one of the four phonon modes
reveals a clear splitting of 3 cm$^{-1}$ as a result of spin-phonon coupling,
the other three optical eigenmodes only show shifts of the eigenfrequencies.
The antiferromagnetic ordering and the concomitant splitting of the phonon mode
can be suppressed in a magnetic field of 7~T. At higher energies we observed a
broad excitation band which is dominated by a two-peak-structure at about
18,000~cm$^{-1}$ and 22,000 cm$^{-1}$, respectively. These energies are in good
agreement with the expected spin-allowed crystal-field transitions of the
Cr$^{3+}$ ions. The unexpected strength of these transitions with $d-d$
character is attributed to a considerable hybridization of the selenium $p$
with the chromium $d$ orbitals.}



\titlerunning{Optical properties of ZnCr$_2$Se$_4$}
\authorrunning{T. Rudolf {\it et al.}}
\maketitle

\section{Introduction}

The $B$-site magnetic ions in the normal spinels $AB_2X_4$ form a corner
sharing tetrahedral network, the well known pyrochlore lattice, which is
strongly frustrated. Specifically in Cr spinels geometrical frustration and
bond frustration compete and produce a complex phase diagram as function of
lattice constant or Curie-Weiss (CW) temperature~\cite{rudolf9.2007}. The
oxides with the smallest lattice constants are dominated by direct
antiferromagnetic (AFM) exchange between neighboring Cr ions and hence are
strongly geometrically frustrated. With increase of the Cr-Cr separation in the
sulfides and selenides, the 90$^{\circ}$ ferromagnetic (FM) Cr-$X$-Cr exchange
gains increasing importance and finally constitutes a FM ground state in
CdCr$_2$S$_4$, CdCr$_2$Se$_4$ and HgCr$_2$Se$_4$. Theoretical calculations of
these exchange couplings have been recently performed within the Local Spin
Density Approximation (LSDA)~\cite{yaresko77.2008}. The majority of the
chromium spinels undergoes AFM order close to 10~K, despite the fact that the
CW temperatures range from -400~K to 140~K~\cite{rudolf9.2007,baltzer151.1966}.
In the AFM systems the magnetic ordering is accompanied by a structural
transition from the cubic normal spinel structure to either tetragonal or
orthorhombic symmetry. The origin of this symmetry reduction has been
attributed to a spin-driven Jahn-Teller
effect~\cite{yamashita85.2000,tchernyshyov88.2002}.

In the case of ZnCr$_2$Se$_4$ the room temperature cubic symmetry with space
group $Fd\bar{3}m$ ($^7O_h$) is lost at $T_{\textrm{N}}=21$~K and the system
was found to become tetragonal with space group $I4_1/amd$ ($^{19}D_{4h}$)
\cite{akimitsu44.1978}. Recently, a minute orthorhombic distortion was reported
leading to the assignment of space group $Fddd$ ($^{24}D_{2h}$)
\cite{hidaka236.2003a}. The magnetic susceptibility can be described by a
\emph{ferromagnetic} CW temperature of $\Theta_{\textrm{CW}}=90$~K and an
effective chromium moment, which is very close to the spin-only value of
Cr$^{3+}$ in an octahedral environment ($S=3/2$)~\cite{hemberger98.2007}. The
spin structure below $T_{\textrm{N}}$ is characterized by ferromagnetic (001)
planes with a turn angle of the spin direction of 42$^{\circ}$ between
neighboring planes~\cite{akimitsu44.1978,hidaka236.2003a}. There are recent
reports on large magnetostriction and on negative thermal
expansion~\cite{hemberger98.2007}, as well as on
multiferroicity~\cite{murakawa77.2008,siratori48.1980}.

The symmetry reduction at the magnetic phase transition can hardly be detected
by conventional X-ray diffraction, but manifests clearly in the lattice
dynamics and a splitting of the infrared (IR) active phonon modes below the
magnetic ordering temperature has been reported for several AFM
spinels~\cite{rudolf9.2007,hemberger97.2006,sushkov94.2005,aguilar77.2008}.
From these splittings the spin-phonon coupling was deduced in some
cases~\cite{sushkov94.2005,aguilar77.2008}. Notably, the magnetic transition in
ZnCr$_2$Se$_4$ can be fully suppressed in external magnetic fields of
6.5~T~\cite{hemberger98.2007} and the splitting of the optical phonons
disappears as cubic symmetry is restored~\cite{rudolf75.2007}.

This paper is intended as a detailed and comprehensive analysis of the IR
experiments in ZnCr$_2$Se$_4$ across the magneto-structural transition and in
the presence of an external magnetic field, which was partly described in a
previous brief publication~\cite{rudolf75.2007}. Moreover, we performed
reflectivity measurements up to 28,000~cm$^{-1}$, where electronic transitions
could be observed, which bear the energy scales of local $d-d$ excitations
within the crystal-field $d$ multiplet.

\begin{figure}[]
\centering
\resizebox{0.8\columnwidth}{!}{%
\includegraphics{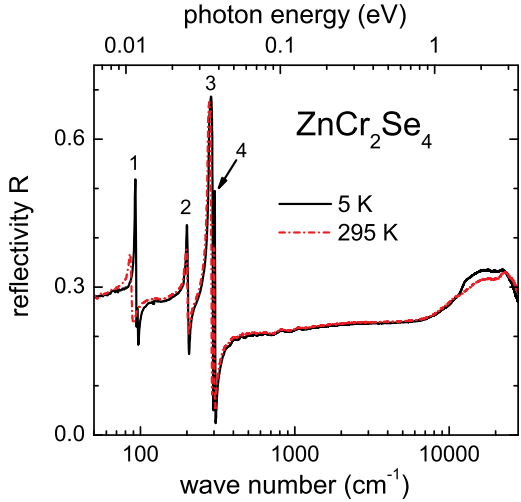}
} \caption{(Color online) Reflectivity vs. wave number for ZnCr$_2$Se$_4$ at
5~K (solid line) and 295~K (dash-dotted line).} \label{fig1}
\end{figure}

\section{Experimental details}

Single crystals with $a = 1.0498(2)$~nm and a selenium fractional coordinate of
$x = 0.260$, which were characterized in earlier
investigations~\cite{hemberger98.2007,rudolf75.2007}, have been used for the
optical experiments. The reflectivity measurements were carried out in the
frequency range from 50 to 28,000~cm$^{-1}$ using the Bruker Fourier-transform
spectrometers IFS 113v and IFS 66v/S, which both are equipped with He bath
cryostats. For the phonon evaluation we analyzed the measured reflectivity
directly utilizing a standard Lorentz model for the complex dielectric function
\begin{equation}
\label{3p}\epsilon(\omega)=\epsilon_{\infty} + \sum_j \frac{\Delta\epsilon\cdot
\omega_{TO,j}^2}{\omega_{TO,j}^2-\omega^2-i\gamma_j\omega},
\end{equation}
with the program RefFIT developed by A. Kuzmenko~\cite{kuzmenko}. For each
phonon mode $j$, the fit parameters are the transverse optical eigenfrequency
$\omega_{TO,j}$, the dielectric strength $\Delta\epsilon$ and the damping
$\gamma_{TO,j}$. The parameter $\epsilon_{\infty}$ results from electronic
polarizabilities only and has been determined from fits up to 5,000~cm$^{-1}$
before the onset of the electronic transitions (see Fig.~\ref{fig1}). We found
that $\epsilon_{\infty} = 8.0(1)$ at room temperature and smoothly decreases to
values close to 7.8(1) at the lowest temperatures. To calculate the dielectric
loss function or the real part of the optical conductivity from the
reflectivity we used the Kramers-Kronig relation with a constant extrapolation
towards low frequencies and a smooth $\omega^{-1.6}$ extrapolation to high
frequencies.

\section{Experimental Results and Discussion}

Figure~\ref{fig1} shows the reflectivity $R$ of ZnCr$_2$Se$_4$ as measured on
single crystalline samples at room temperature (295~K: red dash-dotted line)
and at 5~K (black solid line) for wave numbers ranging from 50 up to
28,000~cm$^{-1}$. At frequencies below 400~cm$^{-1}$ four phonons are detected.
The excitations 3 and 4 at approximately 300~cm$^{-1}$ are very close in energy
and can be distinguished more easily in Fig.~\ref{fig2}. The reflectivity in
the phonon regime is followed by a flat and structureless plateau up to
7,000~cm$^{-1}$ and a subsequent increase due to electronic excitations. Both,
the phonons and the electronic excitations, are obviously temperature dependent
and will be discussed separately in the following.

\subsection{Phonon excitations}

\begin{figure}[b]
\centering
\resizebox{0.95\columnwidth}{!}{%
\includegraphics{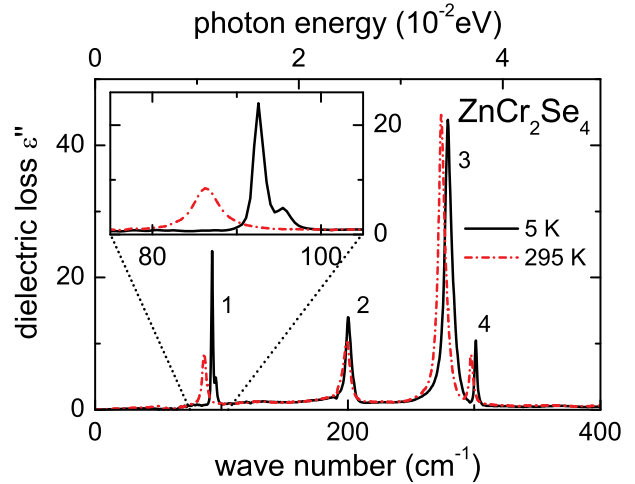}
} \caption{(Color online) Dielectric loss vs. wave number of ZnCr$_2$Se$_4$ for
5~K (solid line) and 295~K (dash-dotted line). The inset shows an enlarged
region around 90~cm$^{-1}$ to document the phonon splitting at low
temperatures.} \label{fig2}
\end{figure}

\begin{figure*}
\centering
\includegraphics[width=18cm]{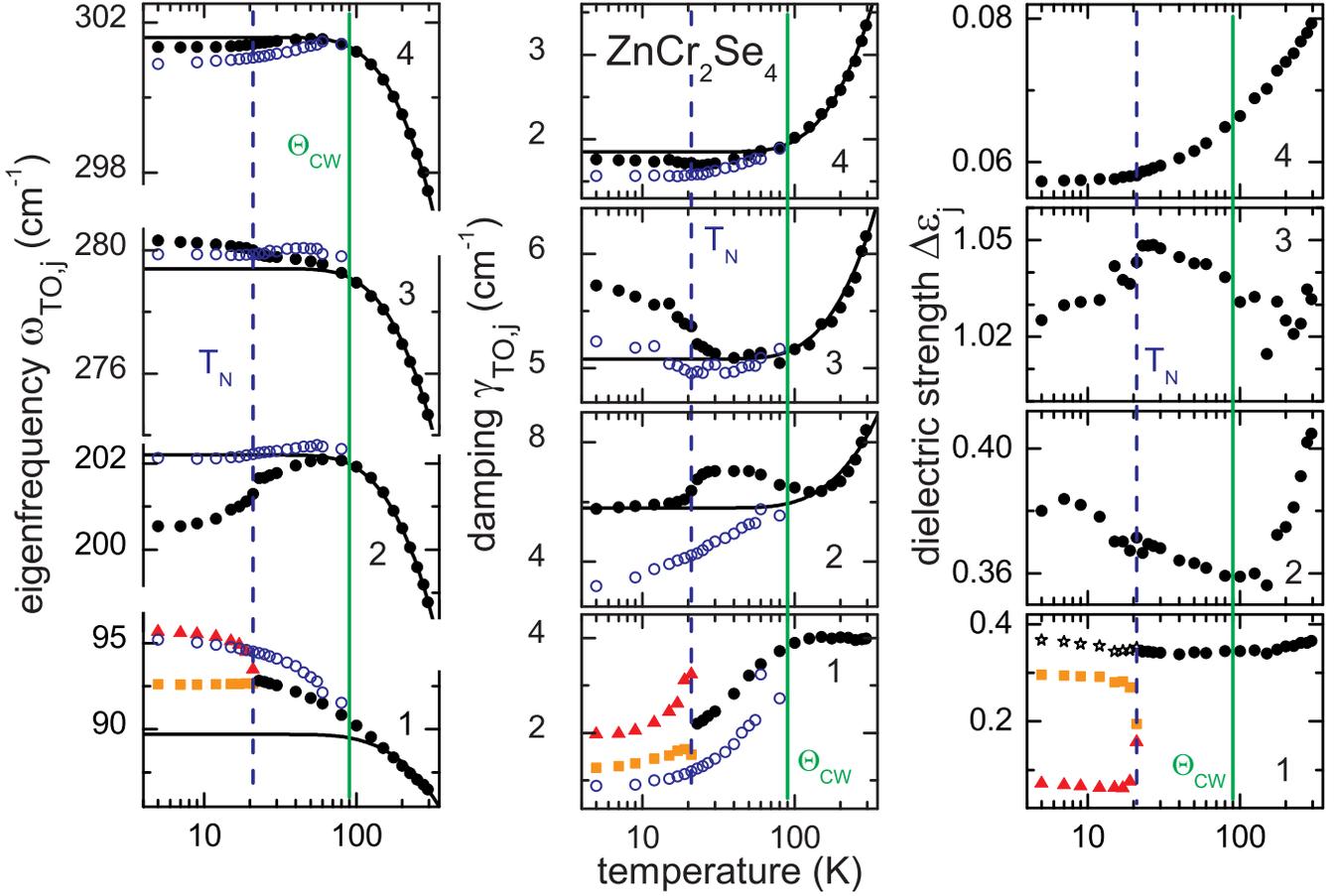}
\caption{\label{fig345}Eigenfrequencies (left frame), damping constants (middle
frame) and dielectric strengths (right frame) of ZnCr$_2$Se$_4$ vs. temperature
for all eigenmodes on semilogarithmic scales. The antiferromagnetic ordering
temperature and the Curie-Weiss temperature are indicated. Closed symbols:
$H=0$~T, open circles: $H=7$~T.}
\end{figure*}

Figure~\ref{fig2} shows the dielectric loss $\epsilon''$ of ZnCr$_2$Se$_4$ in
the phonon regime below 400~cm$^{-1}$, at room temperature and at 5~K, well
below the N\'{e}el temperature ($T_{\textrm{N}} = 21$~K). At 295~K four modes
located approximately at 85, 200, 275 and 300~cm$^{-1}$ can be identified,
corresponding to the four triply-degenerate IR-active $T_{1u}(j)$ modes ($j=1,
2, 3, 4$) expected from a normal mode analysis~\cite{rousseau10.1981}:
\begin{align*}
\Gamma = & \; 4 T_{1u}                              & \text{(IR active)}\\
         & +  A_{1g} + E_{1g} + 3 T_{2g}            & \text{(Raman active)}\\
         & + 2 A_{2u} + 2 E_{u} + T_{1g} + 2 T_{2u} & \text{(silent)}
\end{align*}
Upon cooling shifts in the eigenfrequencies are detected. When entering the
magnetically ordered regime at $T_{\textrm{N}}$ mode $T_{1u}(1)$ reveals a
clear splitting into two modes (see inset of Fig.~\ref{fig2}).

The detailed temperature dependence of the transversal eigenfrequencies and the
corresponding damping constants are shown in the first two columns of
Fig.~\ref{fig345}, without magnetic field and in a field of 7~T. In the third
column we show the temperature dependence of the dielectric strength. The
relevant magnetic temperature scales given by $T_{\textrm{N}}=21$~K and the
ferromagnetic CW temperature $\Theta_{\textrm{CW}}=90$~K are indicated to
reveal a possible influence of ferromagnetic fluctuations and magnetic ordering
on the phonon properties.

Starting with the temperature dependence of the transversal eigenfrequencies in
the left frame of Fig.~\ref{fig345}, one can see that the shift is dominated by
canonical anharmonic effects for $T > 100$~K (black solid lines), i.e. it
originates from phonon-phonon interactions: all phonon modes harden on
decreasing temperature. To get an estimate of the anharmonic effect, we
followed a simple approach, outlined in Ref.~\cite{rudolf76.2007}, where only a
prefactor and an average Debye temperature enter into the fit of the
high-temperature data. The Debye temperature was calculated from the IR active
phonon modes and was kept constant at $\Theta_{\textrm{D}}=308$~K for all fits
shown in Fig.~\ref{fig345}. However, below the temperature scale of
$\Theta_{\textrm{CW}}=90$~K this continuous increase is interrupted. Mode~2 and
4 start to decrease, while modes~1 and 3 increase even stronger than the purely
anharmonic processes suggest. Evidently, mode~4 does not change at
$T_{\textrm{N}}=21$~K, mode~1 splits, and mode~2 and 3 show a shift that might
be considered to mimic the behavior of magnetic sublattice-order parameters.

\begin{table*}
\caption{Table of low temperature structures, observed and expected IR active
phonon modes, and the size of the phonon splitting of different Cr spinels
$A$Cr$_2X_4$.} \label{tab1}
\begin{tabular}{lccccccc}
\hline\noalign{\smallskip}
Compound & $\Theta_{\textrm{CW}}$ & $T_{\textrm{N}}$ & space group & IR-modes    & size of splitting in cm$^{-1}$ for            & new modes at            & Refs.\\
         &                        &       & for $T<T_{\textrm{N}}$ & obs. (exp.) & the $T_{1u}$ mode ($j$) and $T<T_{\textrm{N}}$  & freq. in cm$^{-1}$\\
\noalign{\smallskip}\hline\noalign{\smallskip}
ZnCr$_2$O$_4$  & -398 & 12.5 & $I\bar{4}m2$ & 5 (15)  & 11 (2)$$ & 553$$  & \cite{lee19.2007},\cite{sushkov94.2005},\cite{rudolf9.2007}\\
ZnCr$_2$S$_4$  & 7.9  & 15   & $I4_1/amd$   & 5 (10)  & 4 (3) & & \cite{krimmel.tbp}\\
               &      & 8    & $Imma$       & 12 (20) & 5 (1), 3 (2), 4 (3), 3 (4) & 130, 262, 350, 381 & \cite{krimmel.tbp}\\
ZnCr$_2$Se$_4$ & 90   & 21   & $I4_1/amd$   & 5 (10)  & 3 (1) & & \cite{akimitsu44.1978}\\
HgCr$_2$S$_4$  & 140  & 22   & $Fd\bar{3}m$   & 4 (4)   & 0          & - & \cite{krimmel.tbp}\\
\noalign{\smallskip}\hline
\end{tabular}
\end{table*}

Early on, K. Wakamura proposed a phenomenological model of how AFM and FM
exchange interactions lead to a positive or negative shift of the phonon
eigenfrequencies~\cite{wakamura}, but both experimental and theoretical work
has shown that the coupling between lattice dynamics and exchange couplings is
more intricate~\cite{rudolf9.2007,fennie96.2006}. Furthermore, the significant
differences in the temperature dependence of the damping constants of the
respective phonon modes can not easily be explained along existing
models~\cite{wesselinowa8.1996}.

To sort out the influence of AFM interactions on each mode, we look at the
temperature dependence of the modes in an applied magnetic field of 7~T. At
this field the AFM transition is suppressed and the system remains cubic. It
becomes clear that $T_{1u}(4)$ does not change significantly upon the
suppression of AFM fluctuations and is already dominated by FM interactions in
zero field. Note that also the damping constant of this mode is well described
by anharmonic effects (black solid line) in the whole temperature range. The
dielectric strength decreases towards lower temperatures without any anomalies
at $\Theta_{\textrm{CW}}$ or $T_{\textrm{N}}$.

The positive shift of $T_{1u}(3)$ in zero magnetic field and below
$T_{\textrm{N}}$ seems to originate from the AFM ordering only, as a magnetic
field of 7~T leads to a suppression of that slight positive shift. The damping
of this mode is well described by anharmonic effects with an additional strong
increase below $T_{\textrm{N}}$ which can be suppressed in a magnetic field of
7~T. For decreasing temperatures $\Delta\epsilon$ seems to increase in the
range $T_{\textrm{N}} < T < \Theta_{\textrm{CW}}$ and decreases below
$T_{\textrm{N}}$.

The order-parameter-like negative shift of mode $T_{1u}(2)$  below
$T_{\textrm{N}}$ also disappears in a field of 7~T, relating this shift to the
AFM ordering. The damping of this mode is dominated by anharmonicity with an
additional contribution appearing only between $T_{\textrm{N}} < T <
\Theta_{\textrm{CW}}$. In a magnetic field of 7~T the damping of mode~2
continuously decreases towards lower temperatures without any anomalies. The
dielectric strength decreases between room temperature and
$\Theta_{\textrm{CW}}$ and increases towards lower temperatures for $T <
\Theta_{\textrm{CW}}$.

Obviously, the splitting is absent for mode $T_{1u}(1)$ in 7~T and the
eigenfrequency continuously increases towards lower temperatures. The damping
constants decrease with decreasing temperatures both above and below the
magnetic phase transition. The application of 7~T leads to a somewhat smaller
damping with a similar temperature dependence as in zero field.
$\Delta\epsilon$ of mode~1 and its continuation as a sum of the dielectric
strengths of the split modes behaves similar to mode~2. The dielectric
strengths of the split modes at 5~K yield a ratio of approximately 1:4.

It is important to note that the behavior in 7~T of all modes except
$T_{1u}(2)$ is very similar to the behavior of HgCr$_2$S$_4$ which still orders
antiferromagnetically but does not show a splitting, and to the ferromagnet
CdCr$_2$S$_4$~\cite{rudolf76.2007}. For $T_{1u}(2)$ the shift in the latter
compounds becomes even positive below the magnetic ordering.

Unfortunately, it is not straightforward to single out a simple scheme which
relates the competing exchange interactions with the shift and splittings of
the normal modes in AFM spinel systems like ZnCr$_2$Se$_4$, since both AFM and
FM interactions will be modulated by the lattice dynamics.

Concerning the fact that AFM ordering in these systems is accompanied by a
structural symmetry reduction, a splitting of the degenerate cubic $T_{1u}(j)$
modes and the appearance of new modes is expected. Considering that the most
common structural distortion for spinels occurring below the N\'{e}el temperature
is tetragonal with space group $I4_1/amd$ ($^{19}D_{4h}$)
\cite{akimitsu44.1978,hidaka236.2003a,radaelli7.2005,reehuis35.2003} one finds
the following normal modes~\cite{rousseau10.1981}:
\begin{align*}
\Gamma = & \; 4A_{2u} + 6E_{u}                    & \text{(IR active)}\\
         & + 2A_{1g} + 3B_{1g} + B_{2g} + 4E_g    & \text{(Raman active)}\\
         & + 2A_{1u} + A_{2g} + 2B_{1u} + 4B_{2u} & \text{(silent)}
\end{align*}
This corresponds to a splitting of the four cubic $T_{1u}$ and the appearance
of two additional $E_{u}$ modes which are silent in cubic symmetry. The size of
the splittings, however, may be beyond the experimental resolution and depend
strongly on the competition between FM and AFM interactions.

In Table~\ref{tab1} we compare the number of expected and observed phonon modes
for the three AFM spinels  ZnCr$_2X_4$ ($X=\textrm{O, S, Se}$) and for AFM
HgCr$_2$S$_4$. These four compounds differ strongly with regard to the
competition of AFM and FM exchange couplings. Starting with ZnCr$_2$Se$_4$ with
a positive CW temperature and strong FM interactions, only the lowest lying
phonon splits by 3~cm$^{-1}$ at the magneto-structural transition. The other
splittings and the new modes are obviously beyond the resolution. Note that
recently a small orthorhombic distortion has been found in ZnCr$_2$Se$_4$,
which results in the assignment of space group $Fddd$ ($^{24}D_{2h}$) below
$T_{\textrm{N}}$~\cite{hidaka236.2003a,hidaka236.2003c}. In this case one would
expect to observe 18 IR-active modes~\cite{rousseau10.1981}. In almost FM
HgCr$_2$S$_4$, where AFM order can be suppressed by an external magnetic field
well below 1~T~\cite{tsurkan73.2006}, not a single phonon mode splits.

In the case of ZnCr$_2$S$_4$, AFM and FM interactions almost compensate each
other leading to a relatively low $\Theta_{\textrm{CW}}=7.9$~K ($\approx
T_{\textrm{N}}$) and the system undergoes two magneto-structural transitions,
first at 15~K to the tetragonal $I4_1/amd$ symmetry and below 8~K to
orthorhombic $Imma$~\cite{krimmel.tbp}. Interestingly, only mode 3 splits at
the first transition, whereas a complete splitting and several new modes are
observed in the vicinity of $T_{\textrm{N2}}=8$~K. The splittings of all
$T_{1u}$ modes are of the same order of magnitude.

The system with the strongest antiferromagnetic interaction, ZnCr$_2$O$_4$,
reportedly undergoes a different tetragonal distortion with space group
$I\bar{4}m2$~\cite{lee19.2007}. With the assignment of the Wyckoff positions of
the atoms still lacking, we assume that in this symmetry Zn, Cr, and O occupy
sites with multiplicities of 4, 8, and 16 (similarly to the situation in
spinels with tetragonal $I4_1/amd$ symmetry). Then one would expect 15
IR-active phonons. Experimentally, however, only a splitting of mode~2 with
$\Delta=11$~cm$^{-1}$ has been observed by A. B. Sushkov and
coworkers~\cite{sushkov94.2005}. Moreover, a small anomaly indicative of a new
mode at about 553~cm$^{-1}$ has been reported~\cite{rudolf9.2007}.

Phenomenologically, this comparison suggests that a weak AFM exchange on the
background of a strong FM one ($\Theta_{\textrm{CW}}=90$~K) favors the
splitting of mode~1 ($X=\textrm{Se}$), while a splitting of other modes is
suppressed. Note that in HgCr$_2$S$_4$ ($\Theta_{\textrm{CW}}=140$~K) no
splitting has been observed at all. When the interactions almost compensate
each other more splittings become observable and are of the same order of
magnitude ($X=\textrm{S}$). With the AFM interaction dominating, mode~2 is
split significantly and, again, the splitting of the other modes is suppressed.
For ZnCr$_2$O$_4$, the splitting of mode~2 was estimated by Fennie and
Rabe~\cite{fennie96.2006}, and it is desirable to perform such calculations
throughout the whole phase diagram to reveal the sensitivity of each single
mode on the competing exchange interactions.

\subsection{Electronic Excitations}

\begin{figure}
\centering
\resizebox{0.95\columnwidth}{!}{%
\includegraphics{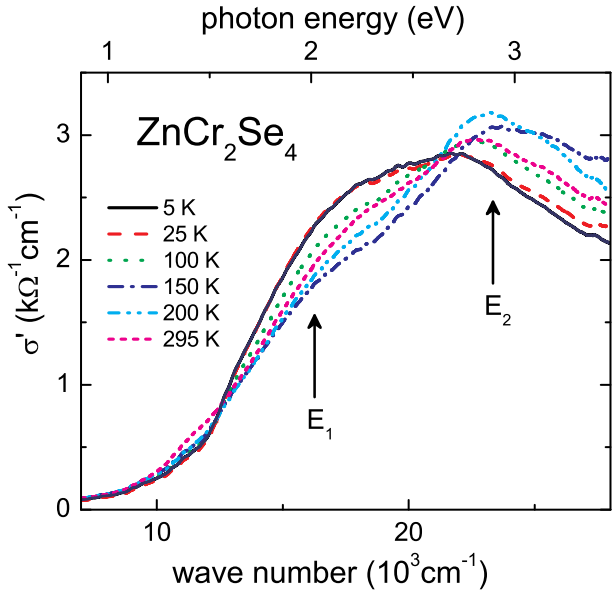}
} \caption{(Color online) Real part of the conductivity vs. wave number of
ZnCr$_2$Se$_4$ between 0.9 and 3.5~eV for a series of temperatures. The
temperatures are given in the figure, the arrows roughly indicate the energies
of the electronic transitions.} \label{fig6}
\end{figure}

Now we turn to discuss the high-energy excitations in ZnCr$_2$Se$_4$. To
analyze the electronic transitions the reflectivity was converted into a
frequency dependent dielectric loss $\epsilon''(\omega)$ and into the real part
of the conductivity $\sigma'(\omega) = \epsilon_0 \omega \epsilon''(\omega)$,
with the dielectric permittivity of vacuum $\epsilon_0$. Both quantities reveal
a two-peak structure (see below) indicating at least two distinct electronic
excitations.

The conductivity (Fig.~\ref{fig6}) is dominated by a strong excitation,
reaching a maximum of $\approx 3,000~\rm{\Omega}^{-1}$cm$^{-1}$, with an onset
at $\approx 10,000$~cm$^{-1}$ and extending up to frequencies
$>28,000$~cm$^{-1}$, beyond our experimentally accessible frequency range. At
room temperature the dielectric loss has been determined from reflectance
spectra in the energy range from 4 to 100~eV by S. Suga {\it et
al.}~\cite{suga25.1982}. Their data show a continuous decrease of
$\epsilon''(\omega)$ between 4~eV and 20~eV. We would like to point out that
similar features are observed in the reflectivity for many chalcogenide spinel
systems, e.g. CdCr$_2$Se$_4$~\cite{zvara12.1979} and
CdCr$_2$S$_4$~\cite{rudolf.tbp}. Historically, the onset of this two-peak
structure has been investigated by absorption measurements, but given the low
transmission of these systems only the very onset of these excitation could be
detected and its changes with temperature became subject to a controversial
discussion about strong and anomalous absorption-edge shifts and the related
magnetic derived changes of the band structure in Cr spinels
\cite{busch23.1966,harbeke17.1966,lehmann1.1970,callen20.1968,white23.1969,kambara28.1970}.

\begin{figure}[h]
\centering
\resizebox{0.95\columnwidth}{!}{%
\includegraphics{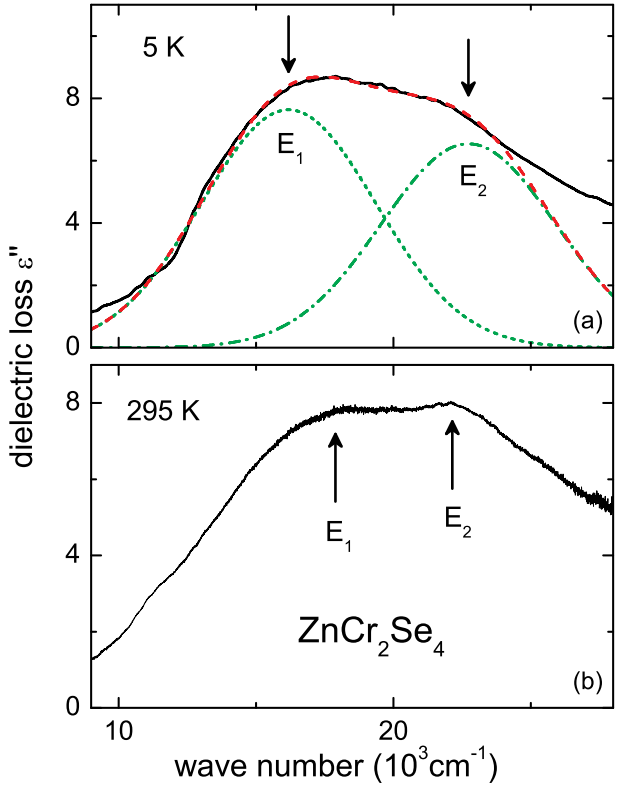}
} \caption{(Color online) Dielectric loss vs. wave number of ZnCr$_2$Se$_4$ at
5~K (upper frame) and 295~K (lower frame). The measurements are indicated by a
thick solid line. The fitting results are given by a dashed line. The two
peaks, which contribute to the total dielectric loss, are indicated.}
\label{fig4}
\end{figure}

\begin{figure}[h]
\centering
\resizebox{0.95\columnwidth}{!}{%
\includegraphics{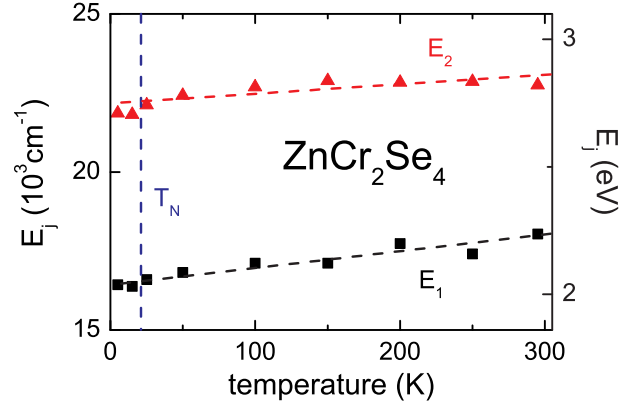}
} \caption{(Color online) Temperature dependence of the transition energies
$E_1$ and $E_2$ of ZnCr$_2$Se$_4$ as determined from the fits as indicated in
Fig.~\ref{fig4}. The dashed lines are drawn to guide the eye.} \label{fig5}
\end{figure}

Early on, there have been suggestions that these absorption edges are not due
to interband transitions, but correspond to excitonic-like onsite $d-d$
excitations within the band
gaps~\cite{berger23.1969,wittekoek7.1969,harbeke8.1970}, while the charge
transfer excitations occur only at higher energies. Indeed, the two maxima at
295~K in $\epsilon''(\omega)$ are located at about 18,000~cm$^{-1}$ and
22,000~cm$^{-1}$, respectively. These frequencies are in good agreement with
typical crystal-field (CF) excitations observed in a large variety of crystals
containing trivalent chromium in an octahedral environment
\cite{wittekoek7.1969,wood39.1963}. To analyze the $d-d$-like excitations in
ZnCr$_2$Se$_4$ in a more quantitative way, we investigated the dielectric loss
vs. wave number (Fig.~\ref{fig4}) by various models. In a first step, we
simulated $\epsilon''(\omega)$ with four Gaussian peaks, taking into account
also spin-forbidden transitions. In this case, the parameters (eigenfrequency
$E_j$, width and intensity) are highly correlated. Even assuming two
spin-allowed transitions with eigenfrequencies $E_1$ and $E_2$ results in an
almost 100\% correlation of width and area of the two peaks. Only constraining
the width of the two peaks to be temperature dependent but equal yields
reliable fits. A prototypical result is shown in Fig.~\ref{fig4}(a) for 5~K.
The temperature dependence of $E_1$ and $E_2$ arising from these fits is shown
in Fig.~\ref{fig5}. The results are very close to those when only the peak
maxima are analyzed. As an example we show the dielectric loss spectrum at
295~K [see Fig.~\ref{fig4}(b)].

In the crystal-field multiplet scheme for Cr$^{3+}$ these two excitations
correspond to the two spin-allowed transitions from the $^4A_{2}$ ground state
to the $^4T_{2}$ and $^4T_{1}$ excited states~\cite{liehr67.1963}. Neglecting
spin-orbit coupling the transition from the ground state to $^4T_{2}$ is
located just at $10~Dq$, where $Dq$ is the ligand-field coulombic parameter.
The Racah parameter $B$ is given by
$B=(2E_1-E_2)(E_2-E_1)/(27E_1-15E_2)$~\cite{wittekoek7.1969}, usually
approximated by $(E_2-E_1)/12$. Hence, the first two excitation energies
directly yield the parameters $Dq$ and $B$. Table~\ref{tab2} lists the values
for ZnCr$_2$Se$_4$ at 295~K in comparison with other Cr containing spinel
systems. Obviously, the two spin allowed CF transitions are approximately at
the same energies for different $A$-site ions and also coincide with the
transitions observed in diluted systems like Cr doped ZnAl$_2$O$_4$. The value
taken from the literature for ZnCr$_2$S$_4$~\cite{golik38.1996} was derived
from absorption and photoconductivity measurements in comparison with
CdCr$_2$S$_4$~\cite{larsen29.1972}. Reflectivity measurements in
CdCr$_2$S$_4$~\cite{rudolf.tbp} show a similar behavior and excitation energies
as in ZnCr$_2$Se$_4$, in contrast to earlier reports. Since the Cr
crystal-field levels seem to be unaffected by changes in the lattice constant
or the competition of FM and AFM interactions for most of these compounds, it
is likely that the assignment of the excitations for ZnCr$_2$S$_4$ and
CdCr$_2$S$_4$ ought to be reconsidered.

\begin{table}
\caption{Electronic $d-d$ transition energies for various spinel systems. The
crystal field parameters $10~Dq$ and $B$ are indicated.} \label{tab2}
\begin{tabular}{lccc}
\hline\noalign{\smallskip}
Compound & $E_1=10 Dq$ & $B$ & Reference\\
\noalign{\smallskip}\hline\noalign{\smallskip}
ZnCr$_2$Se$_4$          & 18041 & 431 & this work\\
ZnAl$_2$O$_4$:Cr$^{3+}$ & 18756 & 695 & \cite{wood48.1968}\\
ZnCr$_2$O$_4$           & 17450 & 525 & \cite{szymczak15.1980}\\
ZnCr$_2$S$_4$           & 14195 & 220 & \cite{golik38.1996}\\
CdCr$_2$S$_4$           & 14195 & 243 & \cite{larsen29.1972}\\
                        & 19900 & 303 & \cite{rudolf.tbp}\\
\noalign{\smallskip}\hline
\end{tabular}
\end{table}

The reason for the aforementioned discussion about the onset of the spin
allowed CF excitation in terms of an optical band gap is due to the very large
oscillator strength of these excitations for the chalcogenides. Usually, local
$d-d$ transitions are parity forbidden, but can e.g. become electric dipole
allowed by an admixture of odd phonons (vibronic coupling). The expected
oscillator strength in such a case is of the order of $10^{-3}-10^{-4}$ and one
would expect an increase of the oscillator strength and a broadening with
temperature~\cite{sugano,deisenhofer101.2008}. Such weak excitations are hardly
visible in reflection and, therefore, these transitions are commonly
investigated by absorption measurements. For ZnCr$_2$O$_4$ strong crystal-field
absorption peaks were reported at frequencies of 17,450~cm$^{-1}$ ($^4A_2
\rightarrow\ ^4T_2$), 22,700~cm$^{-1}$ ($^4A_2 \rightarrow\ ^2T_{2}$) and
23,850~cm$^{-1}$ ($^4A_2 \rightarrow\ ^4T_1$)~\cite{szymczak15.1980}. In the
similar compound CdCr$_2$O$_4$ we could not detect any contribution of the CF
transitions to the reflectivity~\cite{Kant08.tbp}, while in chalcogenide
spinels like ZnCr$_2$Se$_4$ the excitations are strong enough to be considered
as the optical gap. It is not yet understood how the large oscillator strength
can be explained. There is no experimental indication that the parity selection
rule is broken on the octahedral $B$ sites, but it could be due to a combined
effect of phonon-induced symmetry breaking and strong hybridization of the Cr
$d$-orbitals with the $2p/3p/4p$ of the O/S/Se ligands. This strong
hybridization was already derived from photoemission data by M. Taniguchi {\it
et al.}~\cite{taniguchi70.1989}. A similar tendency was recently reported by K.
Ohgushi and coworkers~\cite{ohgushi77.2008} for a variety of oxide and
chalcogenide spinel systems. Recent band structure calculations by A. N.
Yaresko also favor the assumption of $d-d$-like transitions as their
theoretical energy of 2.7~eV is lying close to our experimental
values~\cite{yaresko77.2008}.

The temperature dependence of the two excitations in ZnCr$_2$Se$_4$ is shown in
Fig.~\ref{fig5}. These results correspond to fits using two Gaussian peaks of
equal width as shown in Fig.~\ref{fig4}(a). Both excitation energies exhibit
only a weak temperature dependence. It is important to note that both
excitation energies reveal no significant changes at the AFM ordering
temperature. Obviously, the level splitting is only weakly influenced by the
magnetic ordering.

\section{Summary}

We investigated the optical spectrum of ZnCr$_2$Se$_4$ across the AFM phase
transition at 21~K. The phonon response has been analyzed and discussed in
detail as function of temperature and external magnetic field. The temperature
dependence of eigenfrequency, damping and dielectric strength of each phonon
mode is given. These quantities exhibit a complex pattern, influenced by the
onset of FM fluctuations and by AFM order. The results can only partly be
explained by existing models~\cite{wakamura,wesselinowa8.1996}. The comparison
of the observed splittings and the appearance of new modes for the series of
ZnCr$_2X_4$ with $X=$~O, S, Se reveal a complex response of the phonon modes as
function of the competing exchange interactions. Moreover, a broad two-peak
structure could be observed at higher energies. The excitation energies are in
good agreement with chromium onsite crystal-field excitations. Astonishingly,
the band structure of a fully stoichiometric compound appears to be dominated
by single-ion crys\-tal-field effects. The large oscillator strength seems to
be related with the strong hybridi\-zation effects in chalcogenide spinels
between the chalcogenide $p$ and the chromium $d$ states.

\begin{acknowledgement}
This work partly was supported by the Deutsche Forschungsgemeinschaft through
the German Research Collaboration SFB 484 (University of Augsburg).
\end{acknowledgement}

\subsection*{Note added in proof}

After finalizing this article, evidence was found by synchrotron X-ray
diffraction studies that ZnCr$_2$Se$_4$ exhibits a lower symmetry than
$Fd\bar{3}m$ already in the paramagnetic phase~\cite{krimmel}. Thus, the
possibility arises that the inversion symmetry at the Cr site is broken,
yielding a natural explanation of the huge oscillator strength of the Cr
crystal-field excitations.

\end{document}